\documentclass[parskip=half]{scrartcl}

\usepackage{graphicx}

\usepackage[colorlinks=true]{hyperref}

\usepackage[sortcites]{biblatex}
\usepackage{orcidlink}

\begin{document}
    \title{How does post-quantum cryptography affect Central Bank Digital Currency?}
    \author{%
        Lars Hupel%
        \thanks{Giesecke+Devrient GmbH, Prinzregentenstr. 161, 81677 München, Germany, \href{mailto:lars.hupel@gi-de.com}{lars.hupel@gi-de.com}}%
        \;\;\orcidlink{0000-0002-8442-856X}%
        \and%
        Makan Rafiee%
        \thanks{secunet Security Networks AG, Kurfürstenstr. 58, 45138 Essen, Germany, \href{mailto:makan.rafiee@secunet.com}{makan.rafiee@secunet.com}}%
    }

    \maketitle

    \begin{abstract}
        Central Bank Digital Currency (CBDC) is an emerging trend in digital payments, with the vast majority of central banks around the world researching, piloting, or even operating a digital version of cash. While design choices differ broadly, such as accounts vs. tokens, the wallets are generally protected through cryptographic algorithms that safeguard against double spending and ensure non-repudiation. With the advent of quantum computing, these algorithms are threatened by new attack vectors. To better understand those threats, we conducted a study of typical assets in a CBDC system, describe which ones are most amenable to post-quantum cryptography, and propose an upgrade strategy.
    \end{abstract}

\section{Introduction}

Central Bank Digital Currency (CBDC) is a digital means of payment,
issued by a country's (or region's) central bank, denominated in the
national currency.
Over 130 countries are researching, developing, or piloting a CBDC,
according to the latest data in Atlantic
Council's CBDC tracker.\footnote{\url{https://www.atlanticcouncil.org/cbdctracker/}, accessed
  2023-12-18} An additional 11 have already launched a CBDC. Although
consensus around the precise definition of ``launch'' has yet to
surface, a production system is generally understood to encompass the
following criteria:

\begin{itemize}
\item
  continuous and uninterrupted availability for an indefinite amount of
  time, i.e. no unannounced shutdown,
\item
  real legal tender that can always be exchanged at face value with cash
  and deposit money,
\item
  no system resets, i.e. holdings will remain valid,
\item
  upgrade and maintenance work requires little to no intervention from
  users, except for long-term hardware upgrades, similar to the 2-5 year
  cycle of bank cards and smartphones.
\end{itemize}

Any CBDC operating within this framework will face a multitude of
challenges, including operational, security, and monetary. In this
paper, we are focusing on a core security aspect: cryptography.
In particular, we examine the security requirements of CBDC -- and
more broadly, of comparable digital assets --, the cryptographic
algorithms used to control for these security requirements, and the
implications of quantum computing on these aspects.

\paragraph{Current state}
The current gold standard in asymmetric cryptography, elliptic curves,
are well-understood and widely deployed. Even though multiple sets of
parameters for elliptic curves exist, they all share broadly similar
characteristics regarding key generation, key lengths, and performance.
For example, the ECDSA signature standard with an underlying NIST
secp256r1 curve is available in the vast majority of programming
languages, as well as in dedicated hardware devices such as HSMs or
smart card chips.
This enables excellent performance, which in turn means that secp256r1
enjoys popularity.

\paragraph{Cryptographic evolution}
However, all elliptic curves, including secp256r1, are known not to
withstand the emerging threat of quantum computing. Their security
relies on the discrete logarithm, which cannot be computed efficiently
by classical computers. In other words, with today's methods, it is
computationally infeasible to obtain the private key given only the
public key. An algorithm discovered already in the 90s by Peter W.
Shor can perform this operation efficiently given sufficiently large
quantum computers \cite{Shor1995,Proos2003}.

Consequently, classical asymmetric cryptographic algorithms are at risk,
not only those deployed in CBDCs. While the precise timelines for the
hitherto theoretical turning into practical risk are not yet clear, it
is strongly advisable that post-quantum cryptography be incorporated
into a CBDC's design. The library of employed cryptographic primitives
in digital currencies is huge: extending not only to wallets, but also
to technical components of intermediaries, communication channels, and
management of CBDC supply.

A basic consequence of the criteria for a launched CBDC is
that any replacement or upgrade of cryptographic algorithms needs to be
carried out in a rolling fashion, where new and old algorithm can coexist
for a period of several months to several years, depending on whether the
algorithm is used only in software or also in hardware.

\paragraph{Structure}
The goal of this paper is to, therefore, provide insight into the future
of cryptography as it applies to CBDC. We first give an overview of classical
cryptographic algorithms and what they are used for in the CBDC context
(Section~\ref{classical-cryptographic-algorithms}).
Then, we introduce the threats that quantum computing
poses, as well as post-quantum algorithms that address those threats
(Section~\ref{post-quantum-cryptography}).
Equipped with this, we can then examine the cryptography in
use for CBDC implementations and match them to appropriate algorithms
(Section~\ref{cryptographic-inventory}). Finally, we propose an opinionated framework for rolling
updates, i.e., to put the earlier insights into practice (Section~\ref{rolling-upgrades}).
Over the course of the paper, we will sometimes refer to
cryptocurrencies as examples, as they share many cryptographic aspects
with CBDC, but are often better researched, owing to their longer time
of operation.

\paragraph{Terminology} In general, we define ``wallets'' as containers of
digital assets, and more specifically, CBDC. Wallets can be in hardware
or software form and manage private key material. Typically, ``hardware
wallet'' refers to a physical device that is in the hands of a user.
Naturally, they also require software to operate, but the hardware
aspect refers to the security measures (e.g. secure elements in smart
cards or other embedded devices). In contrast, ``software wallets'' can
mean keys managed by some remote server where users access their funds
through an app that authenticates towards the wallet's operator. Note
that this distinction becomes blurred in case of self-custodial wallets,
where users may use any combination of soft- and/or hardware, including
open source or self-written, to manage their own keys. Following the
majority opinion across central banks, we exclude this from
consideration here.

As for the CBDC model, we follow the standard ``token'' vs. ``account''
distinction (see also Section~\ref{threats-to-digital-assets}), but do not assume either for the
purpose of the paper, except when otherwise noted. When it becomes
necessary in context, the term ``token'' is assumed to refer to key
material that is short-lived.

``Central bank register'' or ``register'' for short, refers to the
centralized storage of transaction and/or token records at a central
bank, which may or may not be implemented as a DLT. In public
cryptocurrencies, this register is typically a proof-of-work or
proof-of-stake blockchain. The register is said to ``validate''
transactions and/or tokens.

\paragraph{Related work}
The Bank for International Settlements (BIS), has analysed risks to
the global financial infrastructure stemming from quantum computing.
In particular, they have built a secure VPN tunnel connection two
central banks~\cite{BIS2023}. In the CBDC context, this is useful for
inter-bank connectivity, but does not apply to wallets.

Ciulei \emph{et al.}~\cite{Ciulei2022} and Allende \emph{et al.}~\cite{Allende2023}
have conducted studies regarding the quantum-resistance of popular
blockchains. The latter paper also provides an ``end-to-end framework for
for post-quantum blockchain networks,'' based on Ethereum~\cite{Allende2023}.
Our paper complements this work with specific considerations pertaining to
hardware wallets, and suggests upgrade strategies for quantum-proofing live
systems.

\section{Classical cryptographic algorithms}\label{classical-cryptographic-algorithms}

When shifting payments from physical cash to digital assets, a multitude
of cryptographic primitives must be employed to control for security
requirements. We will first sketch the security requirements and then
explain which cryptographic primitives are used to satisfy them.

\subsection{Security requirements for cash and digital assets}

When two people engage in a cash transaction, e.g. a customer rendering
banknotes to a merchant, a number of security requirements are already
controlled for.

First, both the payer's and the payee's identities are established
merely through personal trust: Both people trust the other person to be
the rightful sender (or receiver, respectively) of the payment.

Second, authenticity of the banknotes can be established by haptic and
visual properties of the physical objects. In many circumstances, for
example, when the paid amount exceeds a certain threshold, a receiving
party will employ additional authenticity checks of the banknotes: an
ultraviolet lamp, a counting machine, or other devices.

Third, the rightful ownership of the banknotes is proved by merely
demonstrating physical possession.

Fourth, and relatedly to the previous point, the payment is completed
and settled by change of ownership of the banknotes: the cash physically
changes hands.

Fifth, the payer is prevented from double-spending, i.e. using the same
banknotes for two concurring payments: as opposed to data, physical
objects cannot be cloned.

Finally, in a series of multiple of cash payments, it is impossible for
any third party to track the payment patterns.

Applying those security requirements to digital assets, they can be
summarized in more technical terms as follows:

\begin{enumerate}
\item
  Authenticity of sender and receiver
\item
  Authenticity of the asset
\item
  Proof-of-ownership of the asset
\item
  Non-repudiation of transactions
\item
  Prevention of double-spending
\item
  Privacy
\end{enumerate}

For cash, those requirements can be easily derived from its defining
characteristics: its physical nature and it being a bearer instrument.

In contrast, in a digital asset ecosystem, they must be controlled for
with cryptographic primitives. Note that in cryptocurrencies, the first
point is often not guaranteed, whereas in CBDC, some form of identity
will typically be established, at least locally between participants.

\subsection{Classification of cryptographic primitives for digital assets}

To ensure the authenticity of sender and receiver, we employ wallet
certificates. Those wallet certificates include some identifier, and are
derived by a PKI operated by the central bank. The underlying
cryptographic primitive are digital signatures, either based on elliptic
curves or RSA. Wallet signatures can also be used to ensure
non-repudiation of transactions. (See also Section~\ref{wallet-identification} for a more
detailed treatment of wallet identifiers in our framework.)

Furthermore, the wallet certificates can also be used to establish an
end-to-end encrypted channel between the wallets by using a key exchange
algorithm, such as ECDH. The communication channel is then encrypted
with a symmetric cipher, e.g. AES-128 GCM.

The authenticity of the tokens must also be checked using digital
signatures. This comprises the proof-of-ownership from the sender (which
can be checked locally by the receiver), as well as a double-spending
check (which relies on global knowledge). Therefore, a verifying
instance must be involved. In the case of CBDC, this instance can be
centrally operated by the central bank, whereas cryptocurrencies would
typically use a DLT. Since a central bank has complete knowledge of the
tokens in circulation, we refer to its verifying instance as the
``central bank register''.

For CBDC, there is an additional requirement in that the register must
communicate the authenticity of a token to the receiver thereof. (In
blockchains, this is achieved by the receiver monitoring the newly-added
blocks.) Consequently, the register can digitally sign its response
using a well-known certificate (see also Section~\ref{other-long-and-short-lived-keys}).

In summary, both symmetric and asymmetric cryptography must be used to
control for the security requirements in a CBDC ecosystem.

\section{Post-quantum cryptography}\label{post-quantum-cryptography}

In this section, we will explain the new threat model that quantum
computing creates for digital currencies, or more generally, any type of
digital asset. One of the ways to alleviate this problem is to switch to
cryptographic algorithms, some of them novel, that do not suffer from
those threats. Therefore, we give an overview of the top contenders in
NIST's standardization competition for such post-quantum algorithms.
Alternatively or complementary to that, best practices can be applied to
mitigate quantum threats in the short term, which we will discuss here
too.

\subsection{New attack model}\label{new-attack-model}

Recall the overview of classical cryptographic algorithms from above,
with RSA and elliptic curves being the most notable representatives.
Their security relies on the impossibility to factorize a number into
prime factors, and to invert exponentiation in a discrete group
(discrete logarithm), respectively. While the precise details are out of
scope for this paper, we can illustrate the factorization problem using
RSA as an example.

\subsubsection{The factorization problem and quantum advantage}

Assume we generate two large prime numbers $p$ and $q$ at
random. Using classical computers, it is easy to compute their product
$n = pq$. However, computing $p$ and $q$ from
$n$ is hard, with the basic algorithm only being able to try all
possible numbers for $p$, then attempting to compute $q$ as
$n/p$. If the length of the prime factors is doubled, this
naïve algorithm takes four times the time, enabling the security of RSA:
a user merely has to choose large enough prime numbers that
factorization of their product cannot be feasibly achieved with the
amount of computation power an (estimated) attacker has available. The
best known factorization of a composite number has occurred in 2020,
where researchers have computed the prime factors of a 250-digit long
integer (829 bits, with current best practice to use at least 2048-bit
keys in RSA).

More optimized algorithms can run faster than trial-and-error division.
Yet the only known way to provide a substantial speedup to scale to
larger prime factors is to leverage a quantum computer's capability to
first reduce the problem of integer factorization and discrete logarithm
to a periodicity problem and then solve it by applying a Quantum Fourier
transform to solve this new problem. An attacker, therefore, gains
considerable advantage compared to classical computers. This can be
practically implemented using Shor's algorithm, which also has the
capacity to break the discrete logarithm problem (DLP).

The ability to run Shor's algorithm at scale is currently limited
through practical constraints of quantum computers, such as the amount
of qubits available and their reliability. According to Roetteler
\emph{et al.}'s estimation \cite{Roetteler2017}, breaking a 2048-bit RSA key
with Shor's algorithm would require approximately 4000 qubits (excluding
error correction). For 256-bit elliptic curve
keys, the situation is more dire, with only 2330 qubits necessary. The
currently largest-known quantum computer, the IBM ``Osprey'', unveiled
in 2022, sports 433 qubits.
This is a three-fold increase from its 2021 predecessor system
``Eagle'', which had 127 qubits available \cite{IBM2021,IBM2022}.

A 2022 paper with another estimate for breaking the elliptic curve used
in Bitcoin let cryptographer Bruce Schneier to conclude that a practical
threat will arise ``no time soon'' \cite{Schneier2022}.
While it is impossible to estimate precisely when that time will arrive,
it stands to reason that a roadmap with countermeasures should be in
place before that.

\subsubsection{Quantum computing and symmetric cryptography}

Symmetric ciphers and cryptographic hash functions do not rely on
one-way functions but rather on the speed of searching a large key
space.

Grover's algorithm is a quantum algorithm that significantly speeds up
the search for keys in a key space. Fortunately, by doubling the key
sizes of symmetric encryption algorithms to 256 bits and hash algorithms
to 512 bits, even Grover's algorithm does not pose a threat to symmetric
cryptography.

\subsubsection{Threats to digital assets}\label{threats-to-digital-assets}

To make the above discussion concrete, let us consider an example:
assets in digital currencies are typically represented as public-private
keypairs. In the cryptocurrency sphere, the main contenders Bitcoin and
Ethereum both use the secp256k1 elliptic curve as a basis. Knowledge of
the private key corresponding to some digital assets enables the owner
to spend that asset. CBDCs would follow the same or a very similar
model. Shor's algorithm can be used advantageously not just for breaking
RSA keys, but also for breaking elliptic curve keys.

Incoming payments require the sender to know the recipient's public key,
which is often referred to as an ``address''. Conversely, outgoing
payments requires producing a digital signature which uses the private
key, but can be validated just with the knowledge of the public key,
which is typically recorded in a blockchain (or in the case of CBDC, the
register). This allows anyone to send assets to a particular address,
but only the rightful owner to spend from that address. Security of
one's assets relies on keeping the private key confidential.

Using classical computers, keeping the private key confidential, but
allowing the public key to be known or recorded by third parties, does
not threaten security.

However, under a quantum regime, an attacker can use a public key to run
Shor's algorithm, obtaining the corresponding private key. Therefore,
the attacker is equipped with the ability to produce valid digital
signatures. In consequence, stealing the assets.

When sending assets to a particular address, both Bitcoin and Ethereum
do not use the public key directly as an address, but rather a hash of
the public key. As explained in the earlier section, sufficiently long
hashes cannot be easily inverted by quantum computers. While the assets
are at rest, i.e. have been received on an address but not yet spent, a
quantum attacker does not have an advantage over a classical attacker.
But as soon as assets are transferred out, the legitimate owner must
disclose the public key so that the blockchain network (or in the case
of CBDC, the register) can validate the signature. In other words, as
soon as the assets are in transit, they are at risk.

\subsection{Classical mitigation measures}

Not all cryptocurrencies have the same exposure to this threat. Bitcoin,
or more broadly speaking, ``token-based'' (also referred to as ``UTxO'')
digital assets, discourage reuse of the same public-private keypair, and
some variants forbid it outright. The advantage lies within the fact
that an outgoing transaction from an address will always consume the
entirety of the assets associated with that address. For example, if
address $A$ has a balance of 10 €, and its owner wants to send 2 €
to address $B$, the wallet will also move the remaining 8 € to a
freshly-generated address $C$. An attacker breaking the public key
of $A$ will not be able to steal any assets, since they are now
located at $C$, its public key not known to the attacker.

However, if a quantum attacker can outperform the payment network, and
can break $A$'s private key faster than the network or blockchain
can validate and finalize the aforementioned payment, they could
generate a new transaction, moving parts or all of the funds to another
address that is controlled by the attacker.

Ethereum, or more broadly speaking, ``account-based'' digital assets
particularly suffer from this problem, since addresses are routinely
reused. In other words, the first time an outgoing transaction is signed
from an Ethereum address, all assets at rest in that address are
available for subsequent theft from an attacker. According to a 2021
survey by Deloitte, approximately 65\% of all Ether in the public
Ethereum network are stored in addresses with revealed public
keys \cite{Barnes2022}.

To summarize, in the case of public cryptocurrencies, using public key
hashes and frequent address rotation are both feasible, albeit not
perfect, mitigation strategies \cite{Barnes2021}.
CBDCs with a tighter control through the issuing entity and less public
exposure can go further: for example, digital signatures and public keys
that are transferred from a wallet to backend systems or to another
wallet can be additionally protected by end-to-end transit encryption,
such as TLS. Unfortunately, these mitigations are also not perfect and
subject to quantum threats. Therefore, for some aspects of a CBDC, it
becomes necessary to select truly quantum-safe algorithms.

\subsection{NIST competition}

\begin{table}[t]
  \centering
  \begin{tabular}{|l|l|l|}
      \hline
      Type & PKE/KEM & Signature \\\hline
      Lattice-based & CRYSTALS-Kyber & CRYSTALS-Dilithium \\
      && FALCON \\\hline
      Hash-based & n/a & SPHINCS+ \\\hline
  \end{tabular}
  \caption{Overview over the winners of the NIST competition}
  \label{tab:winners}
\end{table}

The National Institute of Standard and Technology (NIST) has a history
of holding competitions to evaluate, select and standardize
cryptographic algorithms. The most prominent of those competitions were
the Advanced Encryption Standard (AES) standardization
process between 1997 and 2000 and the NIST hash
function competition between 2007 and 2012 \cite{NIST2001,NIST2012}.

The competition entails multiple rounds with each round eliminating a
number of submitted algorithms. Research groups are encouraged to submit
algorithms, which are then tested and evaluated by the NIST and the
broader scientific community. NIST specifies evaluation criteria for the
competition and selects the winning algorithm at the end of the process.

During the leading post-quantum cryptography conference PQCrypto in
2016, the NIST announced a new competition for selecting quantum secure
asymmetric cryptographic algorithms \cite{Moody2016}.

While all previous competitions were designated to select exactly one
winning algorithm, such as a symmetric encryption algorithm (AES) or a
flexible hash function (SHA-3), this new competition, however, selects
multiple new asymmetric algorithms as winners. The algorithms are
divided in two subcategories: Public Key Encryption/Key Encapsulation
Method and Signature Scheme.

The reason for selecting multiple algorithms as winners is a result of
the inevitable drawbacks that these new algorithms will bring: Some are
very slow in signature creation, while others have very large key
material. Since there is no one-fits-all solution, algorithms must be
selected for their specific use cases. Selecting multiple winning
algorithms yields greater flexibility in cryptographic agility and
enables implementors to tailor the choice of algorithms to their
specific needs.

Five mathematical one-way functions turned out to be promising for
designing quantum secure algorithms: hash-based, lattice-based,
code-based, multivariate polynomial-based and supersingular elliptic
curve isogeny-based.

In July 2022, the NIST announced the winners of the
competition as well as further algorithms to be considered (Table~\ref{tab:winners}) \cite{NIST2022,NIST2022Round4}.
Lattice-based algorithms and hash-based signatures were the only one-way
functions in the winning algorithms.

The most prominent multivariate and isogeny-based algorithms were both
broken last year in a matter of days and fell out of relevance for
further consideration \cite{Castryck2022,Beullens2022}.

\subsubsection{Winners of the NIST competition}\label{winners-of-the-nist-competition}

\noindent
CRYSTALS-Kyber was the only algorithm selected for PKE/KEM, which does
not yield flexibility in the choice of algorithms. Therefore, in the
following, we will only consider the advantages and drawbacks of the
signature algorithms: CRYSTALS-Dilithium, FALCON and SPHINCS+.

\subsubsection{Comparison of signature algorithms}

In 2020, Sikeridis \emph{et al.} ran a performance study on
quantum-secure signature algorithms to analyze and compare key sizes (Table~\ref{size-comparison}) and
operational run-time (Table~\ref{performance-comparison}) \cite{Sikeridis2020}.

\begin{table}[t]
    \centering
    \small
    \begin{tabular}{|l|rrrrr|}
        \hline
        &&&&& \\

        \begin{minipage}{2cm}\raggedright
        Signature Algorithm
        \end{minipage} & \begin{minipage}{1.8cm}\raggedright
        Public Key Size (Byte)
        \end{minipage} & \begin{minipage}{1.8cm}\raggedright
        Private Key Size (Byte)
        \end{minipage} & \begin{minipage}{1.8cm}\raggedright
        Signature Size (Byte)
        \end{minipage} & \begin{minipage}{2.3cm}\raggedright
        Claimed Classical Security Level
        \end{minipage} & \begin{minipage}{2.3cm}\raggedright
        Claimed PQ Security Level
        \end{minipage}

        \\&&&&&\\\hline

        RSA 3072 & 387 & 384 & 384 & 128 bits & 0 bits \\
        ECDSA 384 & 48 & 48 & 48 & 192 bits & 0 bits \\
        Dilithium II & 1184 & 2800 & 2044 & 100 bits & 103 bits \\
        Dilithium IV & 1760 & 3856 & 3366 & 174 bits & 158 bits \\
        FALCON 512 & 897 & 1281 & 690 & 114 bits & 103 bits \\
        FALCON 1024 & 1793 & 2305 & 1330 & 264 bits & 230 bits \\
        SPHINCS+ & 32 & 64 & 16976 & 128 bits & 64 bits \\\hline
    \end{tabular}
    \caption{Size comparison of signature algorithms, adapted from Sikeridis \emph{et al.} \cite{Sikeridis2020}, Table I, with Specification column omitted (verbatim copy with some data removed for brevity)}
    \label{size-comparison}
\end{table}

\paragraph{Size comparison}
It can be observed that the classical algorithms RSA and ECDSA have
remarkably low key and signature sizes, which cannot be reached by the
quantum secure alternatives. Only SPHINCS+ has much lower key sizes,
which is counteracted by the immensely large signature sizes of over
16,000 bytes and comparatively low levels of post-quantum security
levels.

For approximately the same levels of post-quantum security, Dilithium
and FALCON yield similar public key sizes, albeit much larger than
classical primitives as RSA and ECDSA. FALCON, however, yields
significantly smaller private key and signature sizes than Dilithium.

\begin{table}[t]
    \centering
    \small
    \begin{tabular}{|l|rrrr|}
        \hline
        &&&& \\

        \begin{minipage}{2cm}\raggedright
        Signature Algorithm
        \end{minipage} & \begin{minipage}{2.5cm}\raggedright
        Sign (Mean) (ms)
        \end{minipage} & \begin{minipage}{2.5cm}\raggedright
        Sign (St. Derivation) (ms)
        \end{minipage} & \begin{minipage}{2.5cm}\raggedright
        Verify (Mean) (ms)
        \end{minipage} & \begin{minipage}{2.5cm}\raggedright
        Verify (St. Derivation) (ms)
        \end{minipage}

        \\&&&&\\\hline

        RSA 3072 & 3.19 & 0.023 & 0.06 & 0.001 \\
        ECDSA 384 & 1.32 & 0.012 & 1.05 & 0.020 \\
        Dilithium II & 0.82 & 0.021 & 0.16 & 0.005 \\
        Dilithium IV & 1.25 & 0.021 & 0.30 & 0.012 \\
        FALCON 512 & 5.22 & 0.054 & 0.05 & 0.004 \\
        FALCON 1024 & 11.37 & 0.102 & 0.11 & 0.005 \\
        SPHINCS+ & 93.37 & 0.654 & 3.92 & 0.043 \\\hline
    \end{tabular}
    \caption{Performance comparison of signature algorithms, adapted from Sikeridis \emph{et al.} \cite{Sikeridis2020}, Table II (verbatim copy with some data removed for brevity)}
    \label{performance-comparison}
\end{table}

\paragraph{Performance comparison}
While classical algorithms were strictly better in all size comparisons,
quantum-secure alternatives can have competitive sign and verify
performance. SPHINCS+ is the slowest algorithm and comes with much
longer signature creation and verification times, up to 100x slower
signature creation times than classical ECDSA.

Dilithium and FALCON both have similarly low verify operations, even
outperforming ECDSA by a large margin. Additionally, Dilithium even
outperforms both RSA and ECDSA for signature creating times. Falcons
signature creation performance is significantly higher than Dilithium.

In summary, SPHINCS+ comes with the most drawbacks of both large
signature sizes and slow sign operations but having very small key
sizes. Dilithium and FALCON come with different strengths and
weaknesses: Dilithium has a much better signature creation performance,
while Falcon comes with smaller key and signature sizes. Hence, the
choice of which algorithm to use strongly depends on the concrete size
and requirements and must be tailored to each particular use case.

\section{Cryptographic inventory}\label{cryptographic-inventory}

In this section, we will go into greater detail about the cryptographic
mechanisms employed in a digital asset ecosystem, and how they can be
made fit for post-quantum cryptography. The first point, public-key
infrastructure, is only applicable to a CBDC, whereas the later points
can also be applied to other types of (decentralized) digital assets.

\subsection{Public-key infrastructure}

In order to authenticate participating entities in the CBDC ecosystem,
either to the central bank or to each other, employing certificates is a
natural choice. Certificate owners can either be natural persons, such
as customers of a commercial bank, or abstract entities such as a
centrally managed service or a smart card. Since the trust of customers
to a CBDC is, as the name says, dependent on the customers' trust
towards the central bank, it is also a natural choice to build a Public
Key Infrastructure (PKI) managed by the central bank. Then, the trust in
the authenticity of all entities in the CBDC ecosystem is derived from
trusting one Root Certificate Authority (CA) certificate managed by the
central bank. The central bank can exercise full control over which
entities are eligible to partake in the ecosystem or, alternatively,
delegate the trust by issuing sub CA certificates to trusted partners.

By verifying these certificates and their respective certificate chains,
customers and services could always ensure that communication partners
are trustworthy.

Since certificates usually have long validity periods, cryptographic
agility must be carefully considered and planned, long before the threat
by quantum computers materializes.

While there are different certificate formats, most prominently the
X.509 format, the approaches to tackle quantum agility can be achieved
similarly if the certificate offer extension fields or not (described in
more detail below).

Transitioning to a quantum-secure PKI can be done abruptly, by
deactivating the classical PKI and building a new PKI as soon as the
quantum threat materializes. This approach, however, is in most cases
not applicable nor desirable. Specifically, a CBDC ecosystem must ensure
a constant run-time of the system without any interruptions.

A smooth transition to a quantum-secure PKI is both more desirable and
more challenging. The transition must be initiated in time and would
result in a transition period where both classical and quantum-secure
algorithms are used in certificates simultaneously. These so-called
hybrid certificates can be designed in two ways: composite and
non-composite.

\textbf{Composite certificates} have been under consideration in an IETF
draft \cite{Truskovsky2018}
and describe how certificates with multiple signature algorithms could
look like. There are two options being discussed: with and without
extension fields.

\begin{figure}[t]
    \includegraphics[width=\linewidth]{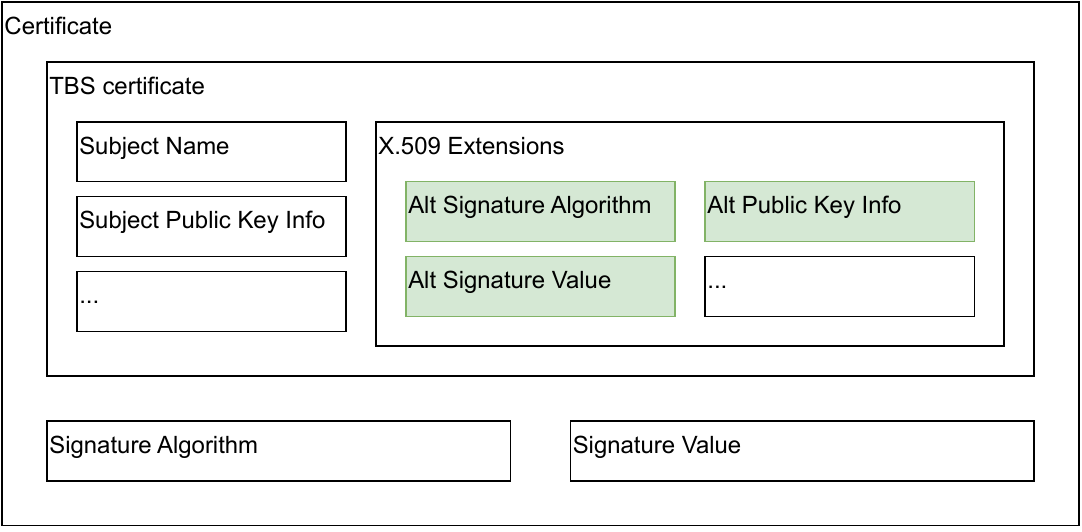}
    \caption{Hybrid certificates (adapted from Vogt and Funke~\cite{Vogt2021} and Truskovsky \emph{et al.}~\cite{Truskovsky2018})}
    \label{composite-certificate}
\end{figure}

In case the certificate format allows for extension fields, as for
example X.509, then a quantum-secure signature algorithm would be
included in a certificate in such an extension field (Figure~\ref{composite-certificate}).

In case extension fields are not existing in the used certificate
format, the classical and quantum-secure algorithm information would
exist in the same field in the certificate; either by concatenation or
by clear separation within the certificate field. In this case,
certificates parsing would need to be adjusted in all relying parties.

In both cases, a transition can be made smoothly. Policy makers could
enforce certificate verifications to entail either only the classical
algorithm, or the quantum-secure algorithm, or both. This policy
adjustment can be made in run-time, and would not require any
interference by the central bank with any circulating certificates.

An inevitable disadvantage of composite certificates is the larger size
of the certificates resulting from the additional key material and
signatures. Particularly bearer-tokens as smart cards could have
difficulties with the resulting larger certificates. An informed choice
of algorithm (see also Section~\ref{winners-of-the-nist-competition}) is necessary.

Alternatively, the idea of \textbf{non-composite certificates} is that
instead of including all the key material in the same certificates,
every entity in the PKI could also be equipped with two separately
issued certificates. These certificates do not need to be issued at the
same time, but can be issued at different points in time. The
certificates must be, however, linked to each other by a reference
mechanism \cite{Becker2022}.

The CSR for the certificate with the quantum-secure certificate would be
signed with the private key corresponding to the classical certificate.
This approach also results in greater flexibility, as the policy makers
could also enforce verification requirements on the relying parties.
Either only the classical-certificate, or only the quantum-secure
certificate, or both verifications must be successful.

\subsection{Algorithm choice in the CBDC ecosystem}

We now discuss suitable algorithms for various types of key material.

\subsubsection{PKI root keys}

The choice of quantum-secure algorithms in a PKI greatly depends on the
requirements arising from which entities create, manage, and verifies
the certificates. For example, a root CA key has highest security
requirements and must be stored securely at the central bank in an
air-gapped HSM. Regular signature operations are not expected and small
private key material is also not necessary. These requirements can,
therefore, be ignored. The public key and signature sizes are important,
however, since all relying parties in the PKI must securely store the
root CA certificate locally to be able to verify all certificate chains.
In the case when hardware is used as a bearer instrument for CBDC, e.g.
a smart card, hybrid certificates for a smooth post-quantum transition
would come with large root CA certificate sizes.

FALCON would be a suitable choice for the root CA signature algorithm,
since the slightly longer signature creation times are outweighed by the
smaller signature sizes.

\subsubsection{Account and wallet keys}

Account and/or wallet keys (depending on the CBDC model) are long-lived:
the key material is created once and is persists for a long time, up to
many years. They are most likely part of the centrally managed PKI to
ensure that only authenticated accounts and wallets partake in the CBDC
ecosystem. That said, the key usage might differ according to the design
specifications of the payment protocols. Account and wallet keys can be
used to establish a secure channel by means of classical ECDH or, for
post-quantum security, CRYSTALS-Kyber, since it is the only selected
PKE/KEM algorithm by NIST. A symmetric AES key is negotiated and derived
by choosing the appropriate key derivation functions (KDFs) and hash
lengths.

For maximum flexibility, it is also a viable option to choose different
algorithms within the same PKI. For example, in the case that wallet
certificates are used to sign CBDC transactions, the expected number of
sign operations is very high in contrast to a root CA certificate.
Furthermore, private key material of a root CA certificate does not have
strict size and performance requirements since it is stored in an HSM,
which are capable of handling large key sizes sufficiently fast.
Conversely, private key material of wallet certificates could be stored
in devices with limited memory.

Therefore, FALCON and Dilithium are viable algorithm choices for root CA
certificates. For end certificates with a high expected number of sign
operations, FALCON is the better choice. SPHINCS+ is not recommended for
either because of the large signature sizes.

\subsubsection{Token keys}

Tokens are keys that are created both at the central bank (minting
process) but also in wallets during regular operations. In contrast to
account and wallet keys, token keys are short-lived and are not part of
any PKI. (See also Section~\ref{token-upgrade-strategies} for a more detailed discussion of cipher
upgrades for tokens.)

Since token keys are created and deleted regularly, a new arising
requirement are short key pair generation times.

While Dilithium has much lower key pair generation times than FALCON or
SPHINCS+ \cite{Raavi2021},
Dilithium signatures are up to three times as large as FALCON signatures
(see above). Therefore, the choice between FALCON or Dilithium for token
key material strongly depends on the concrete design of the payment
protocols and the expected number of key creations and limitations in
memory for token signatures. As mentioned above, SPHINCS+ is not suited
for token keys because of the large signature sizes.

\subsubsection{Other long- and short-lived keys}\label{other-long-and-short-lived-keys}

Apart from the key material discussed above, other keys and certificates
might be used for digital assets. For example, the central bank register
responsible to validate transactions is a crucial component for
security. Transaction must be verified and a receipt must be returned to
the payer or payee to trust and finish the transaction. These receipts
require a trustworthy signature from the register. The certificate used
for receipt signatures has a long validity period with low key size and
key generation time requirements, but must perform an abundance of
signatures to verify transactions. Since Dilithium has much faster
signature creation times than FALCON and SPHINCS+, Dilithium is a good
choice for a register certificate.

\section{Rolling upgrades}\label{rolling-upgrades}

The requirements for a production-grade, widely available CBDC, outlined
in the introductions, imply that the system cannot be stopped for a
coordinated upgrade of cryptographic algorithms. Therefore, any upgrade
roadmap must consider a rolling upgrade strategy. Especially popular in
operations of large-scale distributed systems, a rolling upgrade entails
a step-by-step upgrade of individual components with a specified
timeframe in which old and new versions can coexist and continue to
interact.

Key questions now include:

\begin{itemize}
\item
  How should a component be designed to enable both forward (old system
  understands new instructions) and backward (new system understand old
  instructions) compatibility?
\item
  How long should the timeframe for compatibility be?
\item
  Can the upgrade be coordinated effectively?
\end{itemize}

Despite the similarities between cryptocurrencies and CBDC that we have
alluded to earlier, their answers to these key questions differ
dramatically. Hence, we will focus this section on CBDC.

\subsection{Design decisions that may influence upgrade strategy}

An important dimension in CBDC design is the use of software- and
hardware-based wallets for transactions. In more ambitious designs,
hardware wallets can be used both in online and offline payment
scenarios and exchange money with other hardware wallets, but also
software wallets.

Software-based wallets can come in many shapes, e.g. as a smartphone
app, but also as a database entry managed by a financial intermediary,
such as a commercial bank. In the latter case, upgrades would be
completely transparent to the user.

For hardware-based wallets, the form factor is a core consideration. The
cheapest form factor available is a commodity smart card, like what is
in use today with credit or debit cards. Using CBDC on smart cards
relies on availability of hardware acceleration for cryptographic
algorithms (such as ECDSA with particular curve parameters) to be
feasible. Typically, implementors do not have the ability to choose from
a variety of algorithms.

Another dimension to be considered is whether self-custody is allowed, a
topic that is hotly debated in literature \cite{Narula2023,MortimerLee2023}.
Simply speaking, a CBDC offering self-custody would enable individual
users to provision their own wallets, based on hardware and software
that they deem fit. Central banks, or financial intermediaries that act
as wallet issuers, would have significantly reduced control over the
evolution of such self-custody wallets. For that reason, we exclude them
from further discussions.

\subsection{Wallet identification}\label{wallet-identification}

In cryptocurrencies, addresses are computed from private cryptographic
material. For example, an Ethereum address is computed by hashing an
elliptic curve public key. Wallets would scan the blockchain for any
transaction involving this address. While this avoids the problem of
routing payments to particular entities based on e.g. address prefixes,
like in IBANs, the downside is that the address of the asset is
conflated with the identity to the wallet.

This conflation can be dissolved in CBDCs, where wallet identifiers do
not have to be coupled to the cryptographic keys stored therein. In
fact, a wallet identifier could be akin to a traditional bank account
number, which stays fixed over the lifetime of a bank account, even
though internal implementations details may change. Similarly, a credit
card number would stay the same even after the physical credit card is
replaced.

The reason for this discrepancy is that in CBDC, which should be
accessible to the broad public without in-depth knowledge of
cryptography, it would be poor user experience to change wallet
identifiers if the algorithm changes.

Therefore, we propose that wallet identifiers should not be based on
private cryptographic material, but are randomized upon creation and
remain stable over time (``agnostic identifiers''). Note that such
identifiers are only used for routing in online payment scenarios, i.e.,
for a sending wallet to identify the entity managing the receiving
wallet. Offline payment scenarios do not require routing due to physical
proximity. But more importantly, the central bank register can be made
oblivious to identifiers, since it only concerns itself with digital
signatures from the key material associated with the assets themselves,
therefore avoiding a privacy risk.

The disadvantage of this approach is that agnostic identifiers, as
opposed to a Bitcoin or Ethereum address, cannot be used to deduce which
cryptographic algorithm is supported. This is balanced by the advantage
that agnostic identifiers can support multiple algorithms with full
transparency.

The remainder of this section discusses token upgrade strategies in
general, and hardware concerns based specifically on agnostic
identifiers.

\subsection{Token upgrade strategies}\label{token-upgrade-strategies}

As mentioned earlier, rolling updates are already a routine process in
backend systems. In the case of new cryptographic algorithms, forward
compatibility is much more difficult to achieve than backward
compatibility. As an example, consider adding a data field to a message:
forward compatibility is satisfied if a component can ignore the new
data field and can still process the message. But a new cryptographic
algorithm would need to be available on the component, otherwise the
message cannot be understood.

Instead, focus should be placed on careful backwards compatibility,
coupled with feature detection, if necessary.

Consider a hypothetical upgrade of a cryptographic algorithm, e.g.
replacing ECDSA with FALCON as digital signature scheme for CBDC tokens.
ECDSA and FALCON have different private key formats and lengths,
therefore necessitating a dedicated process.

As a first step, the central bank would equip its register with the new
cryptographic algorithm. In the case of a token-based CBDC, the register
would need to offer a way to convert old-style EC tokens to FALCON
tokens. But since such a CBDC would already support a transaction where
one EC token can be exchanged for another one, simply adding a ``token
version'' field in the token format easily enables this kind of protocol
evolution. The case of hybrid schemes may complicate implementation
details, but can still be achieved in this manner.

This new algorithm becomes available at a certain point in time, with
the central bank mandating a deadline of migrating all tokens. For
simplicity, we use ``new wallet'' and ``old wallet'' to refer to wallets
held in software components that are or are not aware of FALCON yet.
Recall that with agnostic wallet identifiers, the sending wallet would
not be able to deduce the supported key formats of the receiving wallet.

Now, we distinguish between the following cases and subcases for
transactions, and discuss a migration strategy for each:

\begin{enumerate}
\item
  old wallet paying to old wallet

  \begin{enumerate}
  \item
    EC tokens: works unmodified
  \item
    FALCON tokens: would not occur
  \end{enumerate}
\item
  old wallet paying to new wallet

  \begin{enumerate}
  \item
    EC tokens: new wallet needs to be backwards compatible,
    auto-detection possible
  \item
    FALCON tokens: would not occur
  \end{enumerate}
\item
  new wallet paying to old wallet

  \begin{enumerate}
  \item
    EC tokens: works unmodified
  \item
    FALCON tokens: only works if register supports version downgrade
  \end{enumerate}
\item
  new wallet paying to new wallet

  \begin{enumerate}
  \item
    EC tokens: new wallet needs to be backwards compatible,
    auto-detection possible
  \item
    FALCON tokens: works unmodified
  \end{enumerate}
\end{enumerate}

Backwards-compatible auto-detection is trivial to implement, since the
receiving wallet merely needs to check the version of the incoming
token.

The opposite direction, namely detecting if a receiving wallet only
supports an earlier version (case 3) can be solved by imitating
HTTP-style content negotiation: when the sending wallet initiates a
payment, it first contacts the receiving wallet and offers a set of
token versions, which will then be selected by the receiving wallet.

Should a version downgrade (subcase 3b) not be desired or implemented by
the central bank, it is advisable that wallet operators avoid premature
conversion before the deadline for migration (subcases 2a and 4a). This
should be weighed against the possibility to run small trials of a new
cryptographic algorithm to gather more experience, and slowly ramp up
the percentage of wallets that use the new version.

Note that this upgrade strategy can also be applied for other protocol
changes, e.g. increasing the resolution of monetary values (two decimal
digits to four decimal digits).

\subsection{Hardware upgrade strategies}

While token upgrades in hardware wallets follow the same logic as
described in the previous section, there are some additional
considerations relating to their restricted compute power and their
offline capability.

Typical credit or debit cards have expiration times between three to
five years. Therefore, consumers are already familiar with the procedure
of routinely exchanging payment cards. Since simple smart cards are easy
to manufacture and distribute, we anticipate no significant issues
leveraging this strategy to upgrade CBDC hardware wallets.

A problem remains if a CBDC is offline-capable. Some implementations use
deferred communication between hardware wallets and the central bank
register: transactions instantly settle even offline, but wallets keep a
record of digital signatures to upload them to the register and have
them validated at a later point. This may complicate subcase 3b, which
should be avoided in such a setting.

Applied to the above example, this would imply the following timeline:

\begin{itemize}
\item
  At a particular time, FALCON becomes available at the register.
\item
  Soon afterwards, hardware wallets supporting both EC and FALCON become
  available. They never upgrade tokens unless prompted by receiving a
  FALCON token.
\item
  Once all users have obtained new wallets (to be defined as a soft
  deadline), all software wallets upgrade tokens. FALCON tokens start to
  appear in hardware wallets, due to software-to-hardware top-ups.
\item
  Slowly, the hardware wallet ecosystem upgrades, due to
  hardware-to-hardware payments.
\end{itemize}

The hard migration deadline, i.e. the time when the register would no
longer accept EC tokens, would need to be defined to be at least the
soft deadline, plus validity of the new hardware wallets. An additional
safety margin would allow users to exchange their funds in case they let
a wallet sit unused for an extended amount of time, similarly to how the
Eurozone's national banks to this date still allow exchange of local
currency to the Euro.

Perfect enforcement of the aforementioned deadlines is almost impossible
in any manner, since hardware wallets in full offline operation may not
have access to trusted clocks \cite{Bundesbank2021}.

Finally, we will briefly consider point-of-sale (or other payment)
terminals. In the case of online terminals, support for new algorithms
can be added through routine over-the-air updates. For pure offline
terminals, this is naturally not possible.

Therefore, it is advantageous to design the payment protocol between
hardware wallets to be independent of the communication channel and the
terminal. Concretely, the terminal and wallet(s) may have a shared
interface for payment lifecycle (initiation, retry, cancelation), but
the token-level protocol is opaque towards the terminal. This prevents
the need for terminal upgrades altogether. On a low level, this can be
achieved by modularizing terminals and allowing merchants to merely
exchange the smart card (or chip) on which the wallet resides.

\section{Conclusion}

We have discussed how the advent of quantum computing affects the kind
of cryptography used in digital assets, more specifically, a central
bank digital currency (CBDC). In particular, the problem that elliptic
curve private keys could be computed by quantum attackers merely by the
disclosure of the corresponding public key is a looming issue that
threatens safe custody of assets. Even worse, it could lead to loss of
funds without any user interactions whatsoever.

While many concerns are shared between CBDC and cryptocurrency, the
universal nature of CBDC poses additional challenges: for example,
upgrades of cryptographic algorithms must be completely transparent to
users as to not hinder adoption. However, given that CBDC is also more
centralized, it stands to reason that central banks can use their more
granular control over the ecosystem to ensure smooth transitions.

We have shown that a digital asset ecosystem comes with a great deal of
key material with a multitude of different requirements. The choice of
the correct quantum-secure algorithm in accordance with NIST
recommendations is, therefore, not straightforward and must be
considered carefully. In order to grant full flexibility for crypto
agility, we proposed not choosing one algorithm for all components, but
instead select the algorithm for each specific component.

Our proposed cryptographic upgrade framework enables just that and can
be applied to a token-based CBDC design. The framework's cornerstones
are the use of agnostic wallet identifiers, decoupling user-facing
addresses from private key material, and a conservative upgrade strategy
relying on auto-detection of wallet capabilities. We expect this
framework to be used for both software and hardware wallets, and
potentially for self-custodial wallets as well.

\paragraph{Acknowledgements}

We thank our colleague Hermann Drexler for fruitful discussions about post-quantum cryptography.
This work has been partially supported by the Federal Ministry of Education and Research (BMBF), Verbundprojekt CONTAIN (13N16582).

\printbibliography

\end{document}